\documentclass[aps,prb,twocolumn]{revtex4}

\usepackage{amsmath}

\usepackage{graphicx}

\begin{document}

\title{The occupation of a box\\
as a toy model for the seismic cycle of a fault}

\author{\'Alvaro Gonz\'alez}
\email{alvaro.gonzalez@unizar.es}
\author{Javier B. G\'omez}
\email{jgomez@unizar.es}
\affiliation{Departamento de Ciencias de la Tierra, \\ Universidad de
Zaragoza. C. Pedro Cerbuna~12. 50009 Zaragoza, Spain}
\author{Amalio F. Pacheco}
\email{amalio@unizar.es} \affiliation{Departamento de F{\'i}sica
Te\'orica and BIFI, \\
Universidad de Zaragoza. C.~Pedro
Cerbuna~12. 50009 Zaragoza, Spain}

\begin{abstract}
We illustrate how a simple statistical model can describe the
quasiperiodic occurrence of large earthquakes. The model idealizes
the loading of elastic energy in a seismic fault by the stochastic
filling of a box. The emptying of the box after it is full is
analogous to the generation of a large earthquake in which the
fault relaxes after having been loaded to its failure threshold.
The duration of the filling process is analogous to the seismic
cycle, the time interval between two successive large earthquakes
in a particular fault. The simplicity of the model enables us to
derive the statistical distribution of its seismic cycle. We use
this distribution to fit the series of earthquakes with magnitude
around 6 that occurred at the Parkfield segment of the San Andreas
fault in California. Using this fit, we estimate the probability
of the next large earthquake at Parkfield and devise a simple
forecasting strategy.\\

\noindent\textit{American Journal of Physics, Vol.~73, No.~10,
October 2005, pp.~946-952. DOI: 10.1119/1.2013310}\\

\end{abstract}

\maketitle

\section{Introduction}\label{sec1}

In reporting the mechanism of the great California earthquake of
1906, Reid\cite{Reid} presented the elastic rebound theory. It
assumes that an earthquake is the result of a sudden relaxation of
elastic strain by rupture along a fault (a rupture surface between
two rock blocks that move past each other). This theory extended
earlier insights into the relation between earthquakes and faults
by other geologists,\cite{Yeats} especially
Gilbert,\cite{Gilbert,Gilbert506} McKay,\cite{McKay} and
Koto.\cite{Koto} Since its formulation, it has been the basis for
interpreting the earthquakes that occur in faults in the Earth's
upper, fragile crust.

According to Reid's theory, elastic energy slowly accumulates on a fault
over a long time after the occurrence of an earthquake, as the rock blocks
on both sides of the fault are strained by tectonic forces. When the strain
is large enough, the system relaxes by fast rupture and/or frictional
sliding along the fault during the next earthquake. The elastic waves
generated by this sudden event are the seismic waves that seismometers
detect.

The tectonic loading and relaxation process of a fault is cyclic.
The seismic cycle is the time interval between two successive
large earthquakes on the same fault, frequently called
characteristic earthquakes.\cite{Schwartz60} If the seismic cycle
were periodic, earthquake prediction would be easy. There is
increasing information about earthquake occurrences in the seismic
record, compiled with historical data and recognition of ancient
large earthquakes on faults.\cite{Yeats} These data show that the
seismic cycle of any given fault is not strictly periodic. The
reason is that the tectonic loading and relaxation of a fault are
complex nonlinear processes.\cite{GeoComplexity} Moreover, faults
occur in topologically complex networks,\cite{Bonnet343} and an
earthquake occurring in a fault influences what occurs in other
faults.\cite{Harris366}

The duration of the seismic cycle is not constant, but follows a
statistical distribution that can be empirically deduced from the earthquake
time series.\cite{Savage540} This distribution, if it were known, could be
used to estimate the probability of the next earthquake. However, it is not
well known, because there are little data (typically less than ten) in the
earthquake time series for any given fault or fault segment.\cite{Savage540}
To use this probabilistic approach, it is convenient to fit the data to a
theoretical statistical distribution.

Especially since the
1970s,\cite{VereJones,Utsu0,Utsu1,Rikitake504,Hagiwara505,Rikitake,VereJones2}
earthquake recurrence is frequently considered as a renewal
process,\cite{Cinlar,Daley} in which the times between successive
events, in this case the large earthquakes in a fault, are assumed
to be independent and identically distributed random variables. In
this interpretation, the expected time of the next event does not
depend on the details of the last event, except the time it
occurred. In combination with elastic rebound theory, the
probability of another earthquake would be low just after a
fault-rupturing earthquake, and would then gradually increase, as
tectonic deformation slowly stresses the fault again. When an
earthquake finally occurs, it resets the renewal process to its
initial state. Several well-known statistical distributions (such
as the gamma,\cite{Utsu497} log-normal,\cite{Nishenko464} and
Weibull\cite{Hagiwara505,Utsu497,Kiremidjian315,Sieh}) have been
used to describe the duration of the seismic cycle and to
calculate the conditional probabilities of future earthquakes.
These distributions also have been used as failure models for
reliability and time-to-failure problems.\cite{Mann}

More recently, many numerical models have been devised for
simulating the tectonic processes occurring on a seismic
fault.\cite{Main296,BenZion493} These models can generate as many
synthetic earthquakes as desired,\cite{Ward542} so the statistical
distribution of the time intervals between them can be fully
characterized.\cite{Robinson608} Two highly idealized models are
the Brownian passage time model,\cite{Matthews} and the minimalist
model.\cite{VazquezPrada250,Vazquez2,Gomez&Pacheco546} Their
seismic cycle distributions have been used as renewal models, to
fit actual earthquake series and estimate future earthquake
probabilities. \cite{Matthews,Gomez&Pacheco546,Gonzalez} They, as
well as the gamma, log-normal and Weibull distributions, provide a
reasonably good fit to the existing
data.\cite{Utsu497,Gomez&Pacheco546,Gonzalez} The renewal models
have been widely applied, particularly in Japan\cite{Imoto472} and
in the United States,\cite{WGCEP270} to estimate the probabilities
of the next large earthquake for particular faults.

This paper aims to explain how a renewal model can fit the series
of seismic cycles in a particular fault, and how it can be used to
estimate the probability of the next large earthquake in the
fault. For this purpose we will use the process of stochastic
occupation of a box to visualize the progressive loading of a
seismic fault. This box model will be used to fit the series of
characteristic earthquakes, with magnitude around 6, which have
occurred on the Parkfield segment of the San Andreas fault in
California.

In Sec.~\ref{sec2} we present the data of the Parkfield series,
and compute its mean, standard deviation and aperiodicity
(coefficient of variation). The presentation of these data is
important for appreciating the design and tuning of the subsequent
model. Section~\ref{sec3} is devoted to a detailed presentation of
the box model. In Sec.~\ref{sec4} the parameters of the model will
be tuned to fit the Parkfield data series. The comparison between
the model and the data is made in Sec.~\ref{sec5}, and the annual
probability of occurrence of the next large shock at Parkfield is
calculated. In Sec.~\ref{sec6} we introduce a simple forecasting
strategy for the box model and illustrate its effectiveness for
the Parkfield sequence. In Sec.~\ref{sec7} we present our
conclusions.

\section{The Parkfield Series}\label{sec2}
The San Andreas fault runs for 1200\,km, from the Gulf of
California (Mexico) to just north of San Francisco, where it
enters the Pacific Ocean. Fortunately, it does not slide or break
in its whole length as a single earthquake. Rather, as for other
long faults, each earthquake ruptures only one or a few sections
(segments) of its length. During the last century and a half,
several earthquakes with magnitude around 6 have occurred along a
35\,km-long segment of the San Andreas fault that crosses the tiny
town of Parkfield, CA. The apparent temporal regularity of this
series has lead to extensive seismic monitoring in the
area.\cite{Bakun,Roeloffs500} Including the most recent event,
this Parkfield series\cite{Bakun,Bakun2} consists of seven shocks
which occurred on 9 January 1857; 2 February 1881; 3 March 1901;
10 March 1922; 8 June 1934; 28 June 1966; and 28 September 2004.
The durations (in years) of the six observed seismic cycles are
$c_1=24.07$, $c_2=20.08$, $c_3=21.02$, $c_4=12.25$, $c_5=32.05$,
and $c_6=38.25$. The mean of this series is
\begin{equation}
m=\frac{1}{6} \sum_{i=1}^6 c_i=24.62\,\mbox{yr},
\end{equation}
and its sample standard deviation (square root of the bias-corrected
sample variance) is
\begin{equation}
s=\left[\frac{1}{6-1}\sum_{i=1}^6 (c_i-m)^2 \right]^{1/2} = 9.25\,
\mbox{yr}.
\end{equation}
The coefficient of variation, or aperiodicity, is
\begin{equation}
\alpha=\frac{s}{m}=0.3759 \, .
\end{equation}

\section{The box model}\label{sec3}

In this section we will introduce a renewal model based on a
simple cellular automaton. Cellular automata models are frequently
used to model earthquakes and other natural
hazards.\cite{Malamud473} These models evolve in discrete time
steps, and consist of a grid of cells, where each cell can be only
in a finite number of states. Each cell's state is updated at each
time step according to rules that usually depend on the state of
the cell or that of its neighbors in the previous time step. For
example, a grid of cells can represent a discretized fault plane,
and the rules can be designed according to certain friction
laws,\cite{BenZion493} and include stress
transfer\cite{Olami,Preston383,GeoComplexity} and the mechanical
effects of fluids.\cite{Miller495} In the simplest
models,\cite{Newman478,VazquezPrada250} these details are ignored:
the model is driven stochastically, there are only two possible
states for each cell, and the earthquakes are generated according
to simplified breaking rules. The model proposed here is of this
last type. It is simple, easy to describe analytically, and
generates a temporal distribution of seismic cycles comparable to
that of an actual fault.

\subsection{The rules}

\begin{figure}[h]
\begin{center}
\includegraphics[width=8cm]{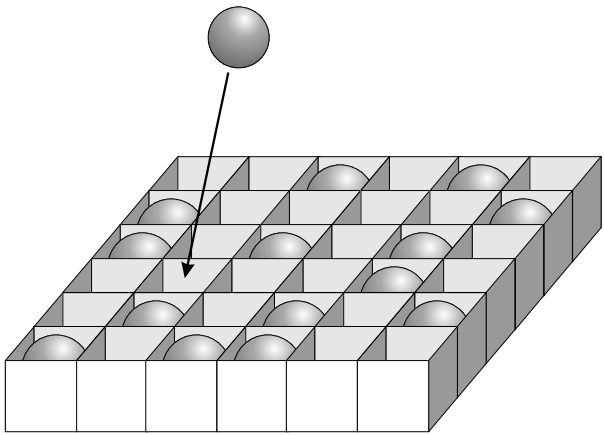}
\caption{\label{fig:boxmodel}Sketch of the box model. Balls are
thrown at random until all the cells of the box are full. Then the
box is emptied and a new cycle starts.}
\end{center}
\end{figure}

Consider an array of $N$ cells. The position of the cells is
irrelevant, but we can assume that they are arranged in the shape
of a box (see Fig.~\ref{fig:boxmodel}). At the beginning of each
cycle, the box is completely empty. At each time step, one ball is
thrown, at random, to one of the cells in the box. That is, each
cell has equal probability, $1/N$, of receiving the ball. If the
cell that is chosen is empty, it will become occupied. If it was
already occupied, the thrown ball is lost. (Thus, each cell can be
either occupied by a ball or empty.) When a new throw completes
the occupation of the $N$ cells of the box, it topples, becoming
completely empty, and a new cycle starts. The time elapsed since
the beginning of each cycle, expressed by the number of thrown
balls, will be called $n$. The duration of the cycles is
statistically distributed according to a discrete distribution
function $P_N(n)$.

The box represents the area of the fault or fault segment, and the
random throwing of balls represents the increase of regional
stress. This randomness is a way of dealing with the complex
stress increase on actual faults. The occupation of a cell by a
ball stands for the elastic strain induced by the stress in a
patch or element of the fault plane. The loss of the balls that
land on already occupied cells mimics stress dissipation on this
plane. The total elastic strain (or conversely the total potential
elastic energy) accumulated in the fault is represented by the
number of occupied cells. This number gradually grows up to the
limit $N$ (analogous to the failure threshold of the fault), and
the toppling of the box represents the occurrence of the
characteristic earthquake in the fault. It is easy to simulate the
evolution of this system with a Monte Carlo approach.

This model is similar to that introduced by Newman and Turcotte
in Ref.~\onlinecite{Newman478}. The difference
is that their model is a square grid of cells in which the
topology is relevant: they consider that the characteristic
earthquake occurs when a percolating cluster\cite{Stauffer} spans
the grid. This cluster happens before the grid is completely
full.

\subsection{Some formulas of the box model}

To describe the box model analytically, it is convenient to recall
some elements of the geometric distribution. Consider the
probability that exactly $x$ independent Bernoulli trials, each
with a probability of success $p$, will be required until the
first success is achieved. The probability that $(x-1)$ failures
will be followed by a success is $(1-p)^{x-1}p$. The resulting
probability function,
\begin{equation}
f(x;p) = (1-p)^{x-1}p,
\end{equation}
is known as the geometric distribution. Its mean and variance are
\begin{equation}
\langle x \rangle = \frac{1}{p}, \qquad
\sigma^{2}=\frac{1-p}{p^2}.
\end{equation}

Now we deal with the box model further. In each cycle, the filling
of the box proceeds sequentially and continues until the $N$th
cell is occupied. Because each of these sequential steps is an
independent process, the mean number of throws to completely fill
the box will be
\begin{equation} \label{eq33}
\langle n \rangle_N=\langle x_1 \rangle_N + \langle x_2 \rangle_N
+ \dots + \langle x_N \rangle_N,
\end{equation}
where $\langle x_i \rangle_N$ is the mean number of throws it
takes to fill the ith cell.

Because the filling of the ith cell is geometrically distributed
with $p_i = (N+1-i)/N$, it follows that
\begin{equation}
\langle x_i \rangle_N=\frac{N}{N+1-i}, \qquad (i=1,2,\dots,N)
\end{equation}
and therefore
\begin{equation}
\label{eq35}
\langle n \rangle_N=1+\sum_{i=2}^{N}\frac{N}{N+1-i}.
\end{equation}

Relations similar to Eqs.~(\ref{eq33}) and (\ref{eq35}) can be written
for the variance of the number of thrown balls to fill the box,
namely
\begin{equation}
\sigma_N^2=\sigma_1^2+\sigma_2^2+\dots=0
+\sum_{i=2}^N\frac{1-\displaystyle\frac{N+1-i}{N}}{\left(
\displaystyle\frac{N+1-i}{N} \right)^2},
\end{equation}
and consequently, the standard deviation is
\begin{equation}
\label{eq37}
\sigma_N=\left[\sum_{i=2}^{N}\frac{N(i-1)}{(N+1-i)^2}\right]^{1/2}.
\end{equation}

The aperiodicity of the series, $\alpha_N$, for a given $N$ is
\begin{equation}
\label{eqalpha}
\alpha_N=\frac{\sigma_N}{\langle n \rangle_N}.
\end{equation}

The mean and the standard deviation of the box model can be
calculated by summing the $N-1$ terms of Eqs.~(\ref{eq35}) and
(\ref{eq37}). For $N \ge 10$, these two equations can be
approximated (with an absolute error $< 0.01$) by their asymptotic
expressions:\cite{asymptotic}
\begin{equation}\label{eq38}
\langle n \rangle_N \xrightarrow[N\to\infty]{} N(C+\ln
N)+\frac{1}{2},
\end{equation}
where $C \simeq 0.5772157$ is Euler's constant, and
\begin{equation}
\label{eq39} \sigma_N \xrightarrow[N\to\infty]{}
N\left[\frac{\pi^2}{6}-\frac{1+C+\ln N}{N}\right]^{1/2},
\end{equation}
and the aperiodicity can be estimated by using Eq.~(\ref{eqalpha})
with Eqs.~(\ref{eq38}) and (\ref{eq39}).

The function $P_N(n)$ is not as easy to obtain as its mean and standard
deviation, and is given by
\begin{equation}
\label{eq311}
P_N(n)=\sum_{j=1}^{N-1} (-1)^{j+1} {N-1 \choose j-1}
\Big(1-\frac{j}{N}\Big)^{n-1},
\end{equation}
and the accumulative probability function, $A_N(n)$:
\begin{eqnarray}
\label{eq312}
A_N(n)&=&\sum_{j=N}^{n}P_N(j)=\\
&=&1-\sum_{j=1}^{N-1}(-1)^{j+1} {N-1 \choose j-1}
\Big(1-\frac{j}{N}\Big)^n\ \frac{N}{j}. \nonumber
\end{eqnarray}
We have deduced Eq.~(\ref{eq311}) by
means of a Markov chain approach analogous to the one used in
Refs.~\onlinecite{VazquezPrada250} and \onlinecite{Vazquez2}. This
derivation is omitted here because of its length.

\section{Fitting the parameters of the box model}\label{sec4}

We will fit the Parkfield series to the accumulative probability
function, Eq.~(\ref{eq312}), using the method of
moments.\cite{Utsu497} Another method that could be used is that
of maximum likelihood.\cite{Utsu497} We have seen in
Sec.~\ref{sec3} that the aperiodicity in the box model depends
only on $N$. Thus, we will choose $N$ for which the aperiodicity
is the closest to that of the Parkfield series, that is, $\alpha
\simeq 0.3759$. The result is $N=11$, for which, from
Eq.~(\ref{eqalpha}), the aperiodicity is $\alpha=0.3752$.

\begin{figure}[!]
\begin{center}
\includegraphics{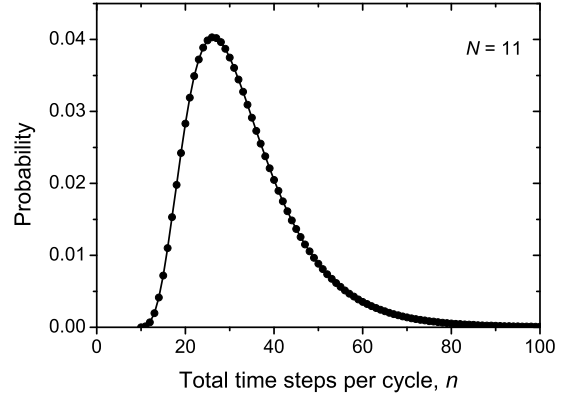}
\caption{\label{fig:P11n} Discrete distribution function for the
duration (in time steps, $n$) of the seismic cycle in the box
model with $N=11$.}
\end{center}
\end{figure}

{}From Eq.~(\ref{eq35}) the mean value of $n$ for $N=11$ is
$\langle n \rangle_{N=11}= 33.22$. Because the actual mean of the
Parkfield series is $m = 24.62$\,yr, one ball throw in the model
is equivalent to $\tau = 24.62\,\mbox{yr}/33.22 = 0.74\,\mbox{yr}
\approx 9$\,months. The discrete distribution function for the
duration of the seismic cycle in a box model with $N=11$,
$P_{11}(n)$ is shown in Fig.~\ref{fig:P11n}.

\begin{figure}[h]
\begin{center}
\includegraphics{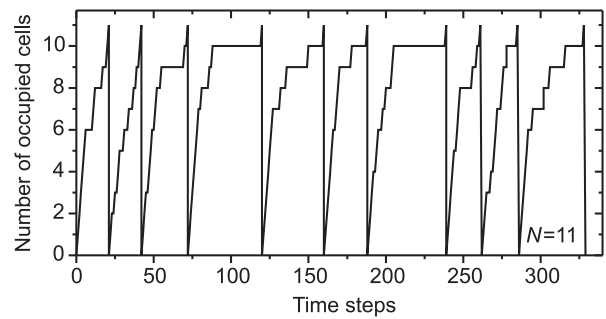}
\caption{\label{fig:cycles} Plot of the number of occupied cells
during ten cycles of a box model with $N=11$.}
\end{center}
\end{figure}

In Fig.~\ref{fig:cycles} we plot the evolution of the number of
occupied cells for ten cycles with $N=11$. Note the fluctuations
in the duration of the cycles, which are consistent with the mean
and the standard deviation of the series. Note also that the
occupation increases rapidly just after a toppling, and then slows
down. This increase is due to the fact that, as a cycle
progresses, there are more occupied cells, and thus it is less
probable for an incoming ball to land on an empty cell. If
$\rho_n$ is the fraction of occupied cells at time step $n$, there
is a probability $1-\rho_n$ for the next ball to be thrown to an
empty cell. Because such a throw would increase $\rho$ by $1/N$,
the mean $\rho$ at step $n+1$ is
\begin{equation}
\label{eqdensity}
\langle \rho_{n+1} \rangle =\langle\rho_n\rangle +
\frac{1}{N}[1-\langle\rho_n\rangle].
\end{equation}
The box is empty at the beginning of the cycle ($\rho_0=0$), so
from Eq.~(\ref{eqdensity}), the mean value of $\rho_n$ is
\begin{equation}
\langle \rho_n \rangle =1-\Big(1-\frac{1}{N} \Big)^n,
\end{equation}
which approaches one asymptotically.

In real faults, the strain also does not increase uniformly during
the seismic cycle. Instead, it follows a trend similar to that of
the number of occupied cells in the box model: the loading rate is
faster just after a large earthquake, and decreases over
time.\cite{Michael548}

The relaxation of a real fault by means of a large earthquake
reduces the stress in the system. Thus a minimum time has to
elapse before the fault accumulates enough stress to produce the
next large earthquake. This effect is called stress
shadow.\cite{Harris366} In the box model there exists a stress
shadow: a characteristic earthquake cannot occur until the $N$th
step in the cycle. According to the box model, this minimum time
for the Parkfield series is $\tau N \simeq 8$\,yr.\\

\section{Earthquake probabilities at Parkfield}\label{sec5}

\begin{figure}[h]
\begin{center}
\includegraphics{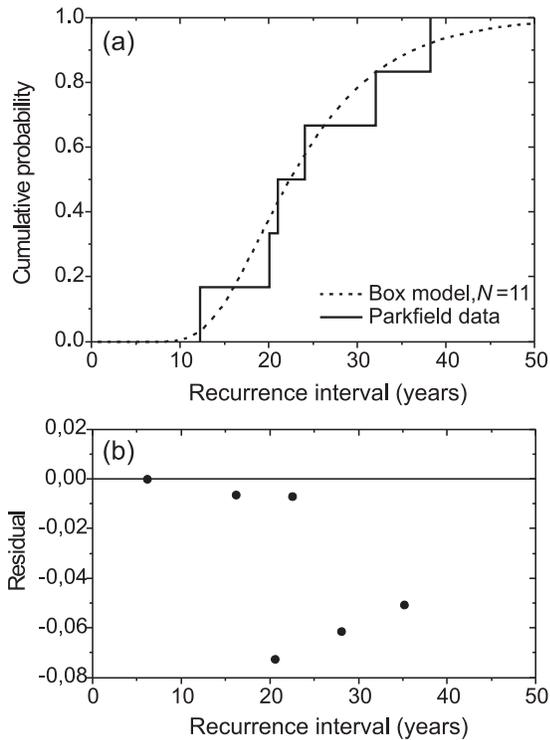}
\caption{\label{fig:fit} (a) Fit of the accumulative distribution
of the box model to the accumulated histogram of the Parkfield
earthquake sequence. (b) Residuals of the fit, evaluated at the
midpoints of the horizontal segments of the accumulated
histogram.}
\end{center}
\end{figure}

We now evaluate the quality of the box model fit for the Parkfield
series and estimate the probability of the next earthquake in this
fault segment. In Fig.~\ref{fig:fit}(a), the empirical
distribution function of the Parkfield series is plotted. It is an
accumulative step function ranging from 0 to 1.0, with a jump 1/6
at each of the six observed recurrence intervals $c_i$. The
accumulated distribution of the box model in Eq.~(\ref{eq312}) for
$N = 11$ with $\tau=0.74$\,yr also is drawn. In
Fig.~\ref{fig:fit}(b), we show the residuals of the fit, which do
not surpass 7.5\%. The equivalent fits to these data, made by
using the renewal models cited in Sec.~\ref{sec1}, give very
similar results.\cite{Gonzalez}

\begin{figure}[h]
\begin{center}
\includegraphics{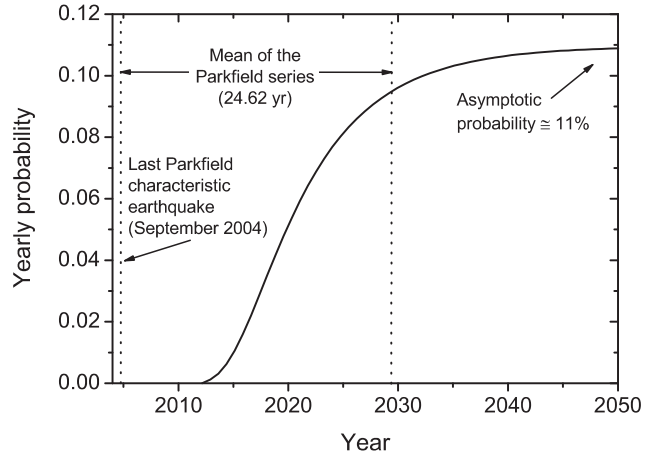}
\caption{\label{fig:yearly} Yearly probability of the next
characteristic earthquake at Parkfield, according to the box
model. }
\end{center}
\end{figure}

Now we calculate the yearly probability of the next
earthquake, that is, the conditional probability of the next
shock occurring in a certain year, given that it has not occurred
previously. For the box model it is
\begin{equation}
\label{eq:conditional}
P_{\tau}(N,n)=\frac{A_N(n+1/\tau)-A_N(n)}{1-A_N(n-1)}.
\end{equation}
Note that $1/\tau$ is the number of time steps of the box model
corresponding to one year. After calculating $P_{\tau}$ from
Eq.~(\ref{eq:conditional}), it is necessary to rescale the
abscissas, $n$, to actual years, $n\tau+t_0$, where $t_0$ is the
calendar year at which the last earthquake occurred ($t_0=2004.75$
for the Parkfield series). In Fig.~\ref{fig:yearly} we plot the
yearly probability for the new cycle at Parkfield according to the
box model. During the first eight years after the last earthquake
at Parkfield (which occurred in September 2004), the box model
indicates that another big shock should not be expected. From that
time on, the probability of the next earthquake increases, tending
to a constant equal to 11\%.

In the seismological literature there is a well-known question
about the yearly probability for a time much longer than the mean
value of the series:\cite{Davis} ``The longer it has been since
the last earthquake, the longer the expected time till the next?''
Sornette and Knopoff\cite{Sornette} have discussed some
statistical distributions that lead to affirmative, negative, or
neutral answers to it. The result shown in Fig.~\ref{fig:yearly}
leads us to conclude that the box model produces a neutral answer.
The reason is that for a long cycle duration (large $n$), the
$P_N(n)$ of the box model decays exponentially, and asymptotically
the box model behaves as a Poisson model, in which the conditional
probability of occurrence of the next earthquake is a constant.

\section{A Simple Forecasting Strategy}\label{sec6}

In earthquake forecasting an ``alarm'' is sometimes turned on when
it is estimated that there is a high probability for a large
earthquake to occur.\cite{KeilisBL} If a large shock takes place
when the alarm is on, the prediction is considered to be a
success. If it takes place when the alarm is off, there has been a
failure to predict. The fraction of errors, $f_e$, is the number
of prediction failures divided by the total number of large
earthquakes. The fraction of alarm time, $f_a$, is the ratio of
the time during which the alarm is on to the total time of
observation. A good strategy of forecasting must produce both
small $f_e$ and $f_a$, because both the prediction failures and
the alarms are costly. Depending on the trade-off between the
costs and benefits of forecasting,\cite{Molchan2} we can try to
minimize a certain loss function, $L$. For simplicity, we will
consider a simple loss function defined as
\begin{equation}
L=f_a+f_e.
\end{equation}

A random guessing strategy (randomly turning the alarm on and off)
will yield $L=1$, a result which can be easily understood. The
alarm will be on, randomly, during a certain fraction of time,
$f_a$. Thus, there will be a probability equal to $f_a$ for it
being on when an earthquake eventually occurs (and a probability
of $1-f_a$ for it being off). The result is that $f_e=1-f_a$. As a
trivial special case, if the alarm is always on ($f_a=1$), then
all the earthquakes are ``forecasted'' ($f_e=0$). Conversely, all
the earthquakes are failures to predict if the alarm is always
off. The random guessing strategy is considered as a baseline, so
a forecasting procedure makes sense only if it gives $f_a+f_e<1$.

\begin{figure}[h]
\begin{center}
\includegraphics{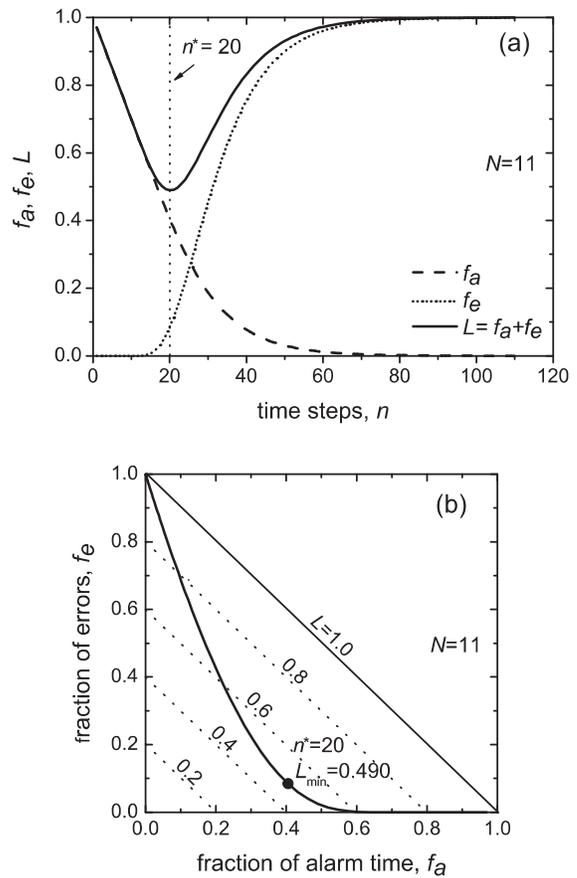}
\end{center}
\caption{\label{fig:forecasting} (a) Fraction of errors ($f_e$),
fraction of alarm time ($f_a$), and loss function ($L=f_a+f_e$) as
a function of $n$ for the forecasting strategy in a box model with
$N=11$. (b) Error diagram for this strategy. Each point on the
curve is the result of using a different value of $n$. The large
dot corresponds to $n^*$, for which the loss function reaches a
minimum. The diagonal lines are isolines of $L$. A random guessing
strategy would render $L=1$.}
\end{figure}

We can use the box model fit to the Parkfield series to implement
a simple earthquake forecasting strategy. The strategy consists in
turning on the alarm at a fixed value of $n$ time steps after the
last big earthquake, and maintaining the alarm on until the next
one.\cite{Newman478,Vazquez2} Then the alarm is turned off, and
the same strategy is repeated, evaluating $f_a$ and $f_e$ for all
the cycles. The result is
\begin{equation}
f_e(n)=\sum_{n'=1}^n P(n'),
\end{equation}
and
\begin{equation}
f_a(n)=\frac{\sum_{n'=n}^\infty P(n')(n'-n)}{\sum_{n'=0}^\infty
P(n')n'}.
\end{equation}
These relations are illustrated in Fig.~\ref{fig:forecasting}(a),
together with $L=f_a+f_e$. For each value of $N$, $L(n)$ has a
minimum at a specific value of $n$, $n^*(N)$. As can be seen in
Fig.~\ref{fig:forecasting}(a), $n^*(11)=20$, for which
\begin{equation}
\label{eq64} f_a(n^*)=0.405, \quad f_e(n^*)=0.085, \quad L(n^*)=
0.490.
\end{equation}
For the Parkfield sequence, $n^*$ corresponds to
\begin{equation}
\tau n^*= 14.8 \, \mbox{yr}.
\end{equation}
If the distribution derived from the box model correctly describes
the recurrence of large earthquakes at Parkfield, the alarm
connected at this time since the beginning of the cycles and
disconnected just after the occurrence of each shock would yield
the results given in Eq.~(\ref{eq64}). Note that this time is
substantially smaller than the mean duration of the cycles,
$m=24.62$\,yr.

The quality of the model-earthquake forecast also can be
understood visually by means of an error diagram,
Fig.~\ref{fig:forecasting}(b), in which $f_e$ is plotted versus
$f_a$.\cite{Molchan2} This kind of plot is similar to the receiver
operating characteristic diagram, used, for example, to test the
success of weather forecasts.\cite{Joliffe}

\section{Summary}\label{sec7}

The generation of large earthquakes in a seismic fault involves
slow loading of elastic strain (or, conversely, elastic energy),
and release through rupture and/or frictional sliding during an
earthquake. The duration of this process is the seismic cycle,
which is repeated indefinitely, leading to a series of recurrent
shocks. We have illustrated this process with a very simple model.
The loading of elastic strain is represented by the stochastic
filling of a box with $N$ cells. The emptying of the box after its
complete filling is analogous to the generation of a large
earthquake, in which the fault relaxes after having being loaded
to its failure threshold. The duration of the filling process is
thus equivalent to the seismic cycle.

The statistical distribution of seismic cycles in the box model
(just as the distributions of the Brownian passage time
model\cite{Matthews} and the minimalist
model\cite{VazquezPrada250,Vazquez2,Gomez&Pacheco546}) can be used
to fit actual earthquake series and to estimate earthquake
probabilities. The conditional probability of the box model has
two interesting features. First, after a large earthquake, there
is a period of stress shadow during which a new large earthquake
cannot occur. Second, from this time on the probability
continuously increases, approaching a constant asymptotic value.
By using a simple forecasting strategy, we have shown that the
earthquakes in the model are predictable to some extent.

We hope that our discussion will be a useful educational tool for
introducing several important geophysical and statistical concepts
to graduate and undergraduate students. It could illustrate how to
make quantitative estimates of a natural phenomenon as popular and
as mystifying as earthquakes.

\begin{acknowledgments}
AFP thanks Jes\'us As\'in, Marisa Rezola, Leandro Moral, and
Jos\'e G. Esteve for useful comments. This research is funded by
the Spanish Ministry of Education and Science, through Project
No.~BFM2002--01798 and Research Grant No.~AP2002-1347 held by
\'AG.
\end{acknowledgments}

\end{document}


{\setlength\arraycolsep{1pt}

\title{Appendix to\\
\
``The occupation of a box as a toy model\\
for the seismic cycle of a fault''\\
\ [\textit{Am.~J.~Phys.}, \bf{73}(10), 946-952]}

\author{\'Alvaro Gonz\'alez}
\email{alvaro.gonzalez@unizar.es}
\author{Javier B. G\'omez}
\email{jgomez@unizar.es} \affiliation{Departamento de Ciencias de
la Tierra, \\ Universidad de Zaragoza. C. Pedro Cerbuna~12. 50009
Zaragoza, Spain}
\author{Amalio F. Pacheco}
\email{amalio@unizar.es} \affiliation{Departamento de F{\'i}sica
Te\'orica and BIFI, \\
Universidad de Zaragoza. C.~Pedro Cerbuna~12. 50009 Zaragoza,
Spain}

\begin{abstract}
This is an informal appendix to the paper ``The occupation of a
box as a toy model for the seismic cycle of a fault''
(\textit{American Journal of Physics}, 73(10), 946-952), where we
illustrated how a simple statistical model can describe the
quasiperiodic occurrence of large earthquakes in a seismic fault.
This appendix describes some proofs that could not be included in
the original
paper because of their length. Namely, we deduce here:\\
(1) the discrete probability distribution for the duration of the
seismic cycle in the model;\\
(2) the asymptotic mean and standard deviation of that
distribution (when the number of cells in the model tends to infinity); and\\
(3) the asymptotic conditional probability in this model (when the
time since the last earthquake tends to infinity).
\end{abstract}
\maketitle

\section{Discrete probability distribution for the duration of the
seismic cycle}

The discrete probability distribution for the duration of the
seismic cycle in the box model was named $P_N(n)$, and written in
Eq.~14 of the original paper \cite{Gonzalez628}.

The box model is a Markov chain \cite{Markov}, and this enables to
deduce $P_N(n)$ by using a technique \cite{Durrett} that we
already applied to the Minimalist Model \cite{VazquezPrada250},
which is also Markovian. A Markov chain is a stochastic process
defined by a discrete random variable $X$ that 1) can only take a
finite number of values, and 2) whose value at the next time step
depends only upon the value at the present time step, being
independent of the way in which the present value arose from its
predecessors. In other words, a Markov chain has no memory: the
evolution of a Markov system at any time depends only on the state
of the system at that time and not on the history of how the state
was achieved.

In a box model with $N$ cells, the state is only determined by the
number of occupied cells, that here will be called $\nu$. The
succession of values of this random variable defines the
stochastic process of the box model. Note that exactly which cells
are occupied is not relevant, but only how many of them are. The
number of stable states in the model is $N$; in each of them $\nu$
takes one value in the set $\{0,1,2,3\dots (N-1)\}$. If $N$ cells
become occupied, the system instantly changes to the empty state.
It does not reside any time step in the state of complete
occupancy, so this is not a stable state.

The value of $\nu$ in the next time step only depends on the value
of $\nu$ in the current time step, so it follows the definition of
a Markov chain. For example, if the system is empty ($\nu=0$), in
the next time step, for sure (with probability equal to 1) it will
move to the state of $\nu=1$. In this second step the fraction of
occupied cells is $1/N$, and the fraction of empty cells is
$(N-1)/N$. So, in the third time step, with probability $(N-1)/N$
another cell will be occupied by a ball ($\nu$ becoming equal to
2), or the model will remain in $\nu=1$ with probability $1/N$
(the probability of the incoming ball landing in the only occupied
cell). In general, for $\nu<N-1$, there is a probability
$(N-\nu)/N$ of moving to $\nu+1$ in the next time step. If
$\nu=N-1$, there is a probability $(N-1)/N$ of moving to $\nu=0$
(passing through $\nu=N$, but not residing any moment there). In
each time step there is a probability $\nu/N$ of residing in the
same state during the next time step.

As for any other Markov chain, for the box model we can define a
transition matrix $\mathbf{M}$, a table that contains all the
transition probabilities of passing, in one time step, from any of
the states of the system to any of the others or to itself. Each
element of the matrix will be denoted in the standard way as
$\mathbf{M}(i,j)$, being $i$ the row (from top to bottom), and $j$
the column (from left to right). Each element gives the
probability of moving from the state $X=i$ in the time step $n$ to
the state $X=j$ in the step $n+1$:
\begin{equation}
\mathbf{M}(i,j)=P(X_{n+1}=j\mid X_n=i).
\end{equation}
The transition matrix of the box model is different for each $N$:
as shown above, the transition probabilities depend on $N$, and
because there are $N$ stable states, the size of the matrix is $N
\times N$. Thus we will denote the matrix for the box model as
$\mathbf{M}_N$. Denoting the occupation state with $\nu$ as above,
the element $\mathbf{M}_N(i,j)$ will be the transition probability
from $\nu=i-1$ to $\nu=j-1$:
\begin{equation}
\mathbf{M}_N(i,j)=P(\nu_{n+1}=j-1\mid\nu_n=i-1).
\end{equation}
The cause for this difference in notation is that $\nu$ ranges
from 0 to $N-1$, while $i$ and $j$ range from 1 to $N$.

Let us now deduce the discrete probability distribution for the
duration of the seismic cycle, using the formalism of Markov
chains. The discrete distribution $P_N(n)$ defines the probability
that, for a box model of $N$ cells, the seismic cycle lasts $n$
time steps. The seismic cycle starts when $\nu=0$, and lasts until
$\nu=0$ again. Except for $N=1$, which is a trivial, special case
of the model, there is no possible transition in one time step
from $\nu=0$ to $\nu=0$ (remember that this impossibility causes
the stress shadow in the model). Speaking more generally, in the
first $n-1$ time steps of the cycle there is no transition to the
state $\nu=0$. Because of this, to calculate $P_N(n)$ we will
first deduce the probability that the system evolves from $\nu=0$
to $\nu=N-1$ in $n-1$ time steps, without having passed through
$\nu=0$ in between. To calculate the probability that the system
evolves from $\nu=N-1$ to $\nu=0$ is simpler. At the beginning of
the $n$-th step the system has $\nu=N-1$. Then a new particle is
added to the only one empty cell, so the occupation becomes
$\nu=N$, but instantly drops to $\nu=0$ at the end of the step.
The transition in the time step is thus from $\nu=N-1$ to $\nu=0$.
The probability for this to happen is $1/N$, the chance for the
incoming particle to land in the only empty cell of the array when
$\nu=N-1$.

Thus, the deduction of $P_N(n)$ proceeds as follows:
\begin{enumerate}
\item Deduce the probabilities of passing between the different
states of the system in one time step. These transition
probabilities will be tabulated in the transition matrix
$\mathbf{M}_N$. \item Remove from $\mathbf{M}_N$ the possibility
of intermediate transitions to $\nu=0$. The resulting matrix will
be called $\mathbf{M}'_N$. \item Calculate the transition
probabilities of passing between the different states in $n-1$
time steps and neglecting the possibility of passing through the
state with $\nu=0$. The result is a new matrix,
$\mathbf{T}_N=\mathbf{M'}_N^{n-1}$. \item One of the elements of
this matrix will indicate the probability of passing from $\nu=0$
to $\nu=N-1$ in $n-1$ time steps without having passed through
$\nu=0$ in between. Multiplying this probability by $1/N$ we will
obtain $P_N(n)$.
\end{enumerate}

Let us proceed in this order. The transition matrices for $N$
equal to 2, 3 and 4 are as follows:\\

\noindent For $N=2$,
\begin{equation}
\mathbf{M}_2=\left(\displaystyle\begin{array}{cc}
0 & 1 \\
\frac{1}{2} & \frac{1}{2}
\end{array} \right)
= \frac{1}{2} \left(\displaystyle\begin{array}{cc}
0 & 2 \\
1 & 1
\end{array} \right);
\end{equation}

\noindent for $N=3$,
\begin{equation}
\mathbf{M}_3=\left(\displaystyle\begin{array}{ccc}
0 & 1 & 0 \\
0 & \frac{1}{3} & \frac{2}{3} \\[.08cm]
\frac{1}{3} & 0 & \frac{2}{3}
\end{array} \right)
= \frac{1}{3} \left( \begin{array}{ccc}
0 & 3 & 0 \\
0 & 1 & 2 \\
1 & 0 & 2
\end{array} \right);
\end{equation}

\noindent and for $N=4$,
\begin{equation}
\mathbf{M}_4=\left( \begin{array}{cccc}
0 & 1 & 0 & 0 \\
0 & \frac{1}{4} & \frac{3}{4} & 0 \\[.08cm]
0 & 0 & \frac{2}{4} & \frac{2}{4} \\[.08cm]
\frac{1}{4} & 0 & 0 & \frac{3}{4}
\end{array} \right)
=\frac{1}{4} \left( \begin{array}{cccc}
0 & 4 & 0 & 0 \\
0 & 1 & 3 & 0 \\
0 & 0 & 2 & 2 \\
1 & 0 & 0 & 3
\end{array} \right).
\end{equation}

\noindent All the elements of these matrices are nonnegative (they
are probabilities) and the sum of all the elements of any row is
always 1. These two are necessary and sufficient properties of
transition matrices of Markov chains. These matrices show evident
regularities, which enable to deduce by inspection that the matrix
for any $N$ is
\begin{equation}
\mathbf{M}_N=\frac{1}{N}\left( \begin{array}{cccccc}
0 & N & 0 & 0 & \dots & 0 \\
0 & 1 & N-1 & 0 & \dots & 0 \\
0 & 0 & 2 & N-2 & \dots & 0 \\
\dots & \dots & \dots & \dots & \dots & \dots \\
0 & 0 & 0 & 0 & N-2 & 2 \\
1 & 0 & 0 & 0 & 0 & N-1
\end{array} \right).
\end{equation}
Note that the matrix multiplied by $1/N$ has only three non-null
diagonals, all of them trivial. The first one is the sequence $N,
N-1, N-2 \dots 2$, the second one is the sequence $0, 1, 2 \dots
N-1$, and the third one is only the bottom left element, which is
always 1.

To calculate $P_N(n)$ the next step (the second one in the list
above) is to prune from this matrix the transitions to $\nu=0$.
The only possible transition to $\nu=0$ is from $\nu=N-1$, and the
probability for this transition is given by the bottom left
element $\mathbf{M}_N(N,1)$. Nullifying this element, the
resulting matrix, $\mathbf{M'}_N$, is particularly simple, because
it has only two trivial, non-null diagonals:
\begin{equation}
\mathbf{M'}_N=\frac{1}{N}\left( \begin{array}{cccccc}
0 & N & 0 & 0 & \dots & 0 \\
0 & 1 & N-1 & 0 & \dots & 0 \\
0 & 0 & 2 & N-2 & \dots & 0 \\
\dots & \dots & \dots & \dots & \dots & \dots \\
0 & 0 & 0 & 0 & N-2 & 2 \\
0 & 0 & 0 & 0 & 0 & N-1
\end{array} \right).
\end{equation}

Now (third step of the list) it is necessary to compute a new
matrix, $\mathbf{T}_N$, which indicates all the transition
probabilities, in $n-1$ time steps, between all the states, with
the restriction that passing from $\nu=N-1$ to $\nu=0$ is
forbidden. In Markov chains, the $m$-step transition probability
matrix is the $m$-th power of the transition matrix
\cite{Durrett}. So the new matrix is
\begin{equation}
\mathbf{T}_N=\mathbf{M'}_N^{n-1}.
\end{equation}
This operation is done through the Jordan decomposition of
$\mathbf{M'}_N$. The element $\mathbf{T}_N(1,N)$ of this matrix is
the transition probability from $\nu=0$ to $\nu=N-1$ in $n-1$ time
steps and with the transition $\nu=N-1 \to \nu=0$ forbidden. As
the probability of passing, in the next time step, from $\nu=N-1$
to $\nu=0$ is $1/N$, $P_N(n)$ is obtained by multiplying that
element by
$1/N$. The results, for $N=2$ and $N=3$ are as follows.\\

\noindent For $N=2$,
\begin{eqnarray}
\frac{1}{2^{n-1}}&=&\frac{2}{2^n}=\frac{1}{2^n}\sum_{j=0}^{2-1}\left[
{2 \choose j} j^{n-1} (-1)^{1-j} (2-j)
\right]=\nonumber\\
&=&\frac{1}{2^n}[0+2];
\end{eqnarray}

\noindent and for $N=3$,
\begin{equation}
\frac{2}{3^{n-1}}(2^{n-2}-1)=\frac{1}{3^n}\sum_{j=0}^{3-1}\left[{3
\choose j} j^{n-1}(-1)^{1-j}(3-j)\right].
\end{equation}

\noindent By inspection, the result for a generic $N$ is
\begin{eqnarray}\label{eq:PN}
P_N(n)&=&\frac{1}{N^n}\sum_{j=0}^{N-1}\left[{N\choose j}j^{n-1}(-1)^{1-j}(N-j)\right]=\nonumber\\
&=&\sum_{j=1}^{N-1} (-1)^{j+1} {N-1 \choose j-1}
\Big(1-\frac{j}{N}\Big)^{n-1}
\end{eqnarray}
(Eq.~14 of the original paper).

\section{Asymptotic mean of $P_N(n)$}

The mean duration of the cycle in the box model was indicated in
Eq.~8 of the original paper: \noindent\begin{eqnarray}
\langle n \rangle_N&=&1+\sum_{i=2}^{N}\frac{N}{N+1-i}=\nonumber\\
&=&
1+\frac{N}{N-1}+\frac{N}{N-2}+\dots+\frac{N}{2}+\frac{N}{1}=\nonumber\\
&=&N\left[\frac{1}{N}+\frac{1}{N-1}+\frac{1}{N-2}+\dots+\frac{1}{2}+\frac{1}{1}\right].
\end{eqnarray}

The asymptotic value of this expression can be obtained
considering that \cite{asymptotic}
\begin{equation}\label{eq:eqEuler}
\sum_{i=1}^{N}\frac{1}{i} \xrightarrow[N\to\infty]{} C+\ln
N+\frac{1}{2N} - 0\left(\frac{1}{N}\right)^2,
\end{equation}
where $C \simeq 0.5772157$ is Euler's constant. Multiplying this
equation by $N$ we obtain the asymptotic mean of $P_N(n)$,
\begin{equation}\label{eq:meanasymp}
\langle n \rangle_N \xrightarrow[N\to\infty]{} N(C+\ln
N)+\frac{1}{2}
\end{equation}
(Eq.~12 of the original paper). The absolute error of this
approximation is $<0.01$ for $N\ge9$.

\section{Asymptotic standard deviation of $P_N(n)$}

The variance of $P_N(n)$ was indicated in Eq.~9 of the original
paper:
\begin{eqnarray}
\sigma_N^2&=&\sum_{i=1}^N\frac{1-\displaystyle\frac{N+1-i}{N}}{\left(\displaystyle\frac{N+1-i}{N}\right)^2}=\nonumber\\
&=&\sum_{i=1}^{N}\frac{1}{\left(\displaystyle\frac{N+1-i}{N}
\right)^2} - \sum_{i=1}^{N}\frac{1}{\displaystyle\frac{N+1-i}{N}}.
\end{eqnarray}
To simplify the sums, we can change the variable to \mbox{$k
\equiv N+1-i$}. Because $i$ ranges from 1 to $N$, $k$ will range
from $N$ to $1$. Then the above equation can be rewritten as
\begin{eqnarray}\label{eq:eqks}
\sigma_N^2&=&\sum_{k=1}^{N}\frac{1}{\left(\displaystyle\frac{k}{N}
\right)^2} -\sum_{k=1}^{N}\frac{1}{\displaystyle\frac{k}{N}}=\sum_{k=1}^{N}\frac{N^2}{k^2}-\sum_{k=1}^{N}\frac{N}{k}=\nonumber\\
&=&N^2\sum_{k=1}^N\frac{1}{k^2}-N\sum_{k=1}^N\frac{1}{k}.
\end{eqnarray}

The first sum in the right-hand side of this expression can be
simplified to
\begin{eqnarray}
\sum_{k=1}^N\frac{1}{k^2} &=& \sum_{k=1}^\infty
\frac{1}{k^2}-\int_{N}^{\infty}
\frac{1}{k^2}\mathrm{d}k=\nonumber\\
&=&\frac{\pi^2}{6}-\left[-\frac{1}{k}\right]_{N}^{\infty}=\frac{\pi^2}{6}-\frac{1}{N}.
\end{eqnarray}
Inserting Eq.~\ref{eq:eqEuler} and this result, Eq.~\ref{eq:eqks}
in the limit of $N\to\infty$ can be written as
\begin{eqnarray}
\sigma_N^2 \xrightarrow[N\to\infty]{}
&&N^2\left(\frac{\pi^2}{6}-\frac{1}{N}\right)-\nonumber\\
&&-N\left[C+\ln N+\frac{1}{2N}-0\left(\frac{1}{N}\right)^2\right]=\nonumber\\
=&&N^2\frac{\pi^2}{6}-N-CN-N\ln
N-\frac{1}{2}=\nonumber\\
=&&N^2\frac{\pi^2}{6}-N(1+C+\ln N)-\frac{1}{2}=\nonumber\\
=&&N^2\left[\frac{\pi^2}{6}-\frac{1+C+\ln
N}{N}-\frac{1}{2N^2}\right].
\end{eqnarray}

The asymptotic standard deviation is the square root of the above
equation,
\begin{equation}
\sigma_N \xrightarrow[N\to\infty]{}
N\left[\frac{\pi^2}{6}-\frac{1+C+\ln
N}{N}-\frac{1}{2N^2}\right]^{1/2}.
\end{equation}
Because $N\to\infty$, the term $-1/2N^2$ can be dropped, so the
equation can be further simplified to
\begin{equation}\label{eq:sigmaasymp}
\sigma_N \xrightarrow[N\to\infty]{}
N\left[\frac{\pi^2}{6}-\frac{1+C+\ln N}{N}\right]^{1/2}
\end{equation}
(Eq.~13 of the original paper). This approximation has an absolute
error $<0.01$ for $N \ge 3$.

The asymptotic aperiodicity, obtained by dividing
Eq.~\ref{eq:sigmaasymp} by Eq.~\ref{eq:meanasymp}, has an absolute
error $<0.0001$ for $N \ge 10$.

\section{Asymptotic conditional probability}
To deduce the asymptotic conditional probability in the box model
we will first consider the asymptotic value of $P_N(n)$ when
$n\to\infty$. This value is the first, largest term in the sum
(when $j=1$ in Eq.~\ref{eq:PN}), namely
\begin{equation}\label{eq:asympPN}
P_N(n) \xrightarrow[n\to\infty]{}
\left(1-\frac{1}{N}\right)^{n-1}=a^{n-1},
\end{equation}
where we have denoted $a\equiv 1-1/N$.

For calculating the asymptotic conditional probability we need to
deduce the value of the cumulative probability distribution,
$A_N(n)$, for that large $n$. This is easier to do by defining the
sum
\begin{equation}\label{eq:ANprima}
A'_N(n)=\sum_{i=n}^{\infty}a^{i-1}=\frac{a^{n-1}}{1-a}.
\end{equation}
Considering that $n$ is large enough, $P_N(n)$ can be replaced by
its asymptotic value (Eq.~\ref{eq:asympPN}), which is the term
summed in $A'_N(n)$. Then it holds that
\begin{equation}
A_N(n-1)=\sum_{i=1}^{n-1}P_N(n) \xrightarrow[n\to\infty]{}
1-A'_N(n),
\end{equation}
so
\begin{equation}\label{eq:ANequality}
A_N(n) \xrightarrow[n\to\infty]{} 1-A'_N(n+1).
\end{equation}

The conditional probability (Eq.~18 of the original paper) is:
\begin{equation}
P_{\tau}(N,n)=\frac{A_N(n+1/\tau)-A_N(n)}{1-A_N(n-1)}.
\end{equation}
\\
\noindent Inserting Eqs.~\ref{eq:ANprima} to \ref{eq:ANequality},
it results that
\begin{eqnarray}
P_{\tau}(N,n) \xrightarrow[n\to\infty]{}&&\frac{1-A'_N(n+1/\tau+1)-[1-A'_N(n+1)]}{A'_N(n)}=\nonumber\\
&&=\frac{A'_N(n+1)-A'_N(n+1/\tau+1)}{A'_N(n)}=\nonumber\\
&&=\frac{\displaystyle\frac{a^{n+1-1}-a^{n+1/\tau+1-1}}{1-a}}{\displaystyle\frac{a^{n-1}}{1-a}}=\nonumber\\
&&=a-a^{1+1/\tau}=a\left(1-a^{1/\tau}\right)=\nonumber\\
&&=\left(1-\frac{1}{N}\right)\left[1-\left(1-\frac{1}{N}\right)^{1/\tau}\right].
\end{eqnarray}
\\
In the original paper we were interested in the yearly conditional
probability for the next large earthquake at Parkfield. In order
to fit the series of previous earthquakes, $N$ was chosen as 11
cells, and one time step corresponded to $\tau=0.74$ years.
Substituting these values in the above equation, the asymptotic
yearly probability when a long time has passed since the last
large earthquake is 0.11 (the value of 11\% cited in the original
paper).

\begin{acknowledgments}
The Spanish Ministry of Education and Science funded this research
by means of the project FIS2005--06237, and the grant AP2002--1347
held by \'A.G.
\end{acknowledgments}


\title{Erratum: ``The occupation of a box as a toy model for the seismic
cycle of a fault'' [Am. J. Phys. 73 (10),  946--952 (2005)]}

\author{\'Alvaro Gonz\'alez}
\email{alvaro.gonzalez@unizar.es}
\author{Javier B.\ G\'omez}
\email{jgomez@unizar.es}
\affiliation{Departamento de Ciencias de
la Tierra, \\ Universidad de Zaragoza. C.\ Pedro Cerbuna 12.\ 50009
Zaragoza, Spain}
\author{Amalio F. Pacheco}
\email{amalio@unizar.es}
\affiliation{Departamento de F{\'i}sica
Te\'orica and BIFI,
Universidad de Zaragoza.\ C.\ Pedro Cerbuna 12.\ 50009 Zaragoza,
Spain}

\begin{abstract}
\noindent\textit{American Journal of Physics, Vol.~75, No.~3,
March 2007, pp.~286. DOI: 10.1119/1.2426356}\\
\end{abstract}

\maketitle

The values of $f_e$ plotted in Fig.~6 and used to estimate $n^*$
were determined by a Monte Carlo simulation of the model
instead of using Eq.~(20). If, at a given time step in the
simulation, the ``alarm'' was sounded and the model earthquake
occurred, the latter was deemed as successfully forecasted. This assumption
is incorrect, and leads to a value of $f_e$ smaller than the true
one in Eq.~(20). Given that $n$ is the number of time steps
before sounding the alarm, if the earthquake occurs
at the $n$th time step, the alarm has still not been sounded, and
the earthquake should be considered a prediction failure. An
earthquake in the box model cannot occur before the $N$th time
step of each cycle, so $f_e=0$ if and only if $n<N$. This
error caused $f_e=0$ also for $n=N$.

\begin{figure}[!t]
\begin{center}
\includegraphics[width=6.5cm]{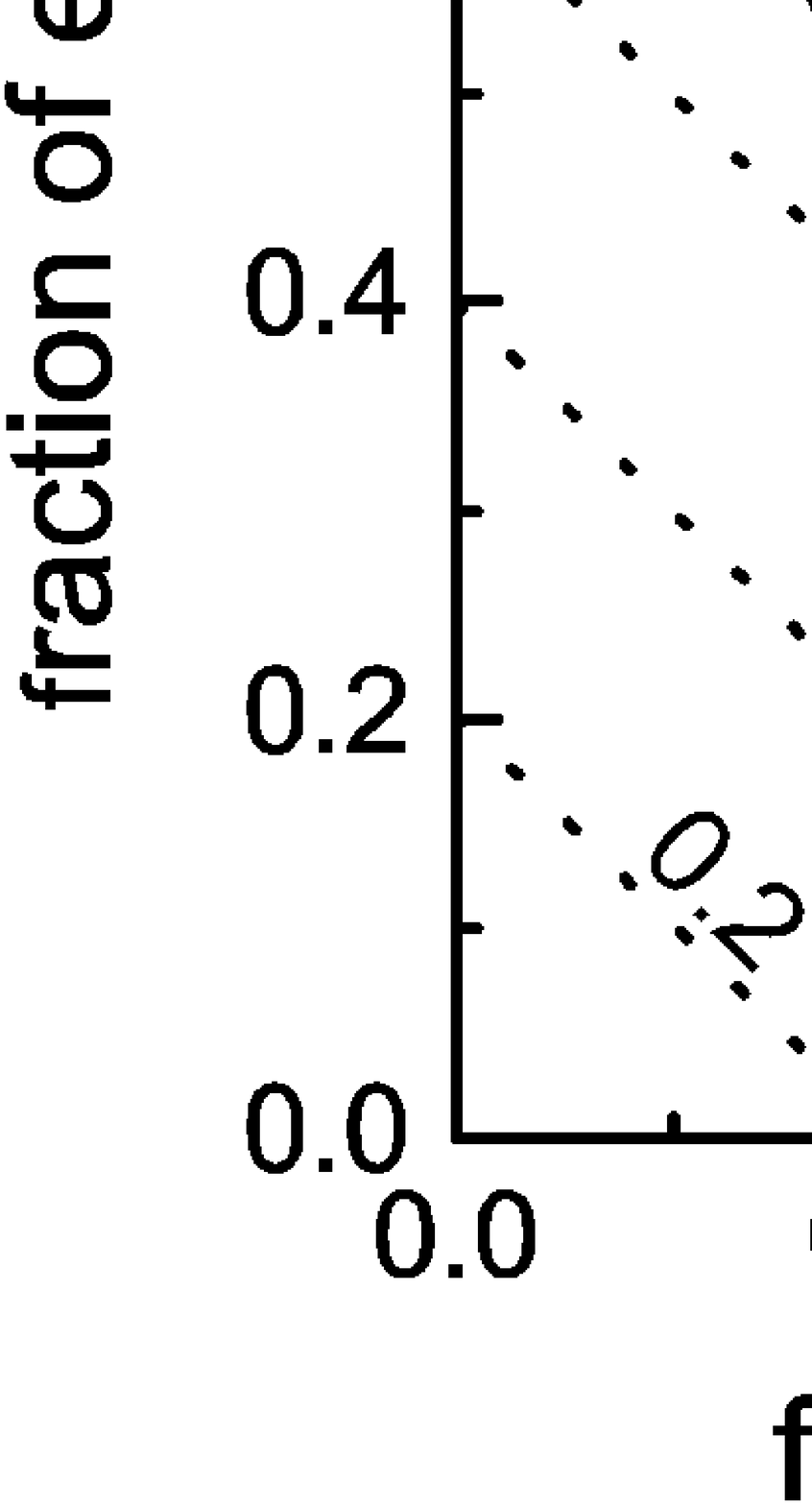}
\caption{Corrected version of Fig.~6 of the original article.}
\end{center}
\end{figure}

We give here a revised version of Fig.~6. The correct values
of $f_e$ and $L=f_a+f_e$ are only slightly higher than those
previously published. This correction changes the value
of $n^*$ (19 time steps, instead of 20). It also modifies the
results in Eq.~(22),
\begin{equation}
f_a(n^*)=0.432, \quad f_e(n^*)=0.084, \quad L(n^*)= 0.516,
\end{equation}
and the value in Eq.~(23) for the Parkfield sequence:
\begin{equation}
\tau n^*= 14.1\,\mbox{yr}.
\end{equation}

We apologize for this error and hope that this note
will serve to clarify the convention for calculating $f_e$ with
discrete probability distributions.